\documentstyle[epsfig,amssymb]{article}
\begin{document}
\title{\large\bf Long-range correlation energies and off-diagonal
interactions for the $\pi$ electronic systems}
\author{\begin{tabular}{c} Hua Zhao \\
\begin{small}
Department of Physics and Institute of Condensed Matter Physics,
College of Mathematics and Physics,
\end{small}
\\
\begin{small}
ChongQing University, ChongQing, 400044, P.R.China
\end{small}
\end{tabular}}

\maketitle \footnote{E-mail:huazhao@cqu.edu.cn}
\begin{abstract}
The long-range correlation energies and the off-diagonal
interactions are studied and a general formula for correlation
energy $E_c$ of the $\pi$ electron systems is given, which is beyond
the nearest-neighbor electron-electron interactions and includes the
off-diagonal interactions. It is found that the effects of the
off-diagonal interactions $W$ and $X$ on the correlation energies
are opposite, but the influence of $X$ on the correlation energies
is much stronger than that of $W$ on the correlation energies, and
the correlation energies decrease with increasing the screening
effect.
\end{abstract}
\vspace*{0.5cm}


function, plasmons \vspace*{0.5cm}

\section{Introduction}

In the previous paper (Ref.[1]), the author deduced a simple
expression for the long-range electronic correlation energies $E_c$
of the $\pi$ electronic systems where only the nearest-neighbor
electron-electron interaction $v$ between two adjacent lattices are
considered, and the formula was applied to the polymer polyacetylene
(PA) and used to calculate the correlation energies and the
calculating results are well consistent with those obtained by other
theoretical calculations including $ab$ $initio$ method[2]. In order
to obtain more general correlation energies formula which includes
the more long-range Coulomb interactions (that is the diagonal
interactions) and also the off-diagonal interactions, the
correlation energy formula in the Ref.[1] should be revised.

As we all have known, besides the diagonal interactions, there are
off-diagonal interactions in the $\pi$ conjugated polymers. The
nearest-neighbor diagonal interactions are the on-site Coulomb
interaction $U$ (Hubbard interaction) and the nearest-neighbor
Coulomb interaction $v$. The nearest-neighbor off-diagonal
interactions have $X$ and $W$, which are so-called, respectively,
the bond-charge interaction and the bond-bond interaction[3]. There
have been many studies about the roles of the off-diagonal
interactions[4]-[8]. There are also the studies about the effects of
the off-diagonal interactions on the excitonic binding energies of
the $\pi$ conjugated polymers [9][10]. It was also known that the
off-diagonal interactions have relationship with the screening from
the $\pi$ electrons in the conjugated polymers[11]. At the different
screening and different conditions, the effects of the off-diagonal
interactions on the physical properties of the electron systems may
have different influences. For instance, $X$ can be used to describe
the possible superconducting states of the organic polymers or the
non-$\pi$ electron systems[12], and $W$ is used to describe the
ferromagnetism in the polymers[13]. Until now, these issues have not
been studied and there is no detailed studies about the contribution
to the correlation energies from the off-diagonal interactions. Thus
the correlation energies from the off-diagonal interactions should
be studied. So, it is necessary to deduce a formula of the
correlation energies where the off-diagonal interactions have been
included explicitly and in the same time the Coulomb interactions
are not limited to the nearest-neighbor interaction $v$.

Therefore, in this study we mainly consider the general situation
where both the long-range electron-electron interactions and the
off-diagonal interactions are considered at the same time as well as
the relation of the screening and the correlation energies.

\section{Theoretical formation }

As the procedure of deducting the desired formula is similar to that
used in the Ref.[1], so here some of the deduction procedure are
omitted and the detailed are mainly referred to the Ref.[1]. We
directly start to our discussion from the following correlation
energy (see the expression (10) in the Ref.[1])
\begin{eqnarray}
 E_c &=&\frac{1}{2}\sum_{\sigma\sigma'}\int d^3r d^3r'
 \rho_{\sigma}(\vec{r})v(\vec{r}-\vec{r'})
 \rho^c_{\sigma\sigma'}(\vec{r},\vec{r'})
 \equiv \frac{1}{2}(I-I_{HF})
 \label{eq:Ec}
\end{eqnarray}
where $v(\vec{r}-\vec{r'})=\frac{e^2}{|\vec{r}-\vec{r'}|}$, the
Coulomb interaction between two charges at $\vec{r}$ and $\vec{r'}$.
In calculation of conjugated polymers, it is parameterized by the
Ohno potential. Here $I$ equals
\begin{eqnarray}
I\equiv \sum_{\sigma\sigma'}\int\int d^3r d^3r'
\rho_{\sigma}(\vec{r})v(\vec{r}-\vec{r'}) \rho_{\sigma'}(\vec{r'})
\widetilde{g}_{\sigma\sigma'}(\vec{r},\vec{r'})
 \label{eq:I}
\end{eqnarray}
and
\begin{eqnarray}
I_{HF}\equiv \sum_{\sigma\sigma'}\int\int d^3r d^3r'
\rho_{\sigma}(\vec{r})v(\vec{r}-\vec{r'}) \rho_{\sigma'}(\vec{r'})
 \widetilde{g}^{HF}_{\sigma\sigma'}(\vec{r},\vec{r'})
 \label{eq:I-HF}
\end{eqnarray}
Here the pair-distribution $\widetilde{g}(\vec{r},\vec{r'})(\equiv
\sum_{\sigma\sigma'}\widetilde{g}_{\sigma\sigma'}(\vec{r},\vec{r'}))$
satisfies the sum rule: $\int
d^3\vec{r'}\rho(\vec{r'})[\tilde{g}(\vec{r},\vec{r'})-1] =-1.$ Using
the relation $\rho_{\sigma}(\vec{r})=
\sum_{ij}^N\rho^{\sigma}_{ij}\phi^*_i(\vec{r})\phi_j(\vec{r})$ and
the sum rule, (\ref{eq:I}) reads
\begin{eqnarray}
I\equiv \frac{1}{2}\sum_{\sigma\sigma'}\sum_{ij,kl}
 \rho^{\sigma}_{ij}\rho^{\sigma'}_{kl}
\int\int d^3r d^3r' \phi^*_i(\vec{r})\phi_j(\vec{r})
v(\vec{r}-\vec{r'}) \phi^*_k(\vec{r'}) \phi_l(\vec{r'})
-\frac{1}{2}\sum_{\sigma}\sum_{ij}\rho^{\sigma}_{ij} \int d^3r
\phi^*_i(\vec{r})v(\vec{r})\phi_j(\vec{r})
 \label{eq:I-3}
\end{eqnarray}
The two-fold integration in the first term in (\ref{eq:I-3}) is a
many-center integral which describes the Coulomb interaction between
two $\pi$ electrons lying, respectively, at the positions $\vec{r}$
with spin $\sigma$ and $\vec{r'}$ with spin $\sigma'$; it is denoted
generally as $V_{ij,kl}$. As for $V_{ij,kl}$, generally it is hard
for us to analytically calculate this many-centered integral easily
and simply. In the CNDO (complete neglect of differential overlap)
approximation[14], $V_{ij,kl}$ reduces to the Coulomb integrals
$V_{ii,kk}$ which is a two-center integral. In the situation of the
conjugated polymers, $V_{ii,kk}$ is further parameterized into the
Ohno potential, that is, $V_{ii,kk}\equiv
V_{i,k}=U/\sqrt{1+((\vec{r}_i-\vec{r}_k)/a_0)^2}$, where $U$ is the
on-site electron-electron interaction[15]. If the overlap integrals
are explicitly considered, the two-electron interaction integral
$V_{ij,kl}$ in (\ref{eq:I-3}) may be approximated as
$US_{ij}S_{kl}/\sqrt{1+
((\vec{r}_{\bar{ij}}-\vec{r}_{\bar{kl}})/a_0)^2}$, where
$S_{ij}$($=\int d^3 r\phi^*_i(\vec{r})\phi_j(\vec{r})$) is the
overlap integral, and $\vec{r}_{\bar{ij}}$ means the position vector
between the site $i$ and the site $j$[16]. When the overlap
integrals are omitted, $S_{ij}$ becomes $\delta_{ij}$, and then
$V_{ij,kl}$ goes back to the above expression $V_{ii,kk}$. After
rearrangement, we obtain
\begin{eqnarray}
I & \equiv &\frac{\lambda}{2}\sum_{\sigma\sigma'}\sum^{i\neq
k}_{i,k}
 V_{ii,kk}\rho^{\sigma}_{ii}\rho^{\sigma'}_{kk}
+ \frac{\lambda}{2}\sum_{\sigma\sigma'}\sum^{k\neq l}_{i,kl}
 V_{ii,kl}\rho^{\sigma}_{ii}\rho^{\sigma'}_{kl}
+ \frac{\lambda}{2}\sum_{\sigma\sigma'}\sum^{i\neq j,k\neq
l}_{ij,kl}
 V_{ij,kl}\rho^{\sigma}_{ij}\rho^{\sigma'}_{kl}
 \nonumber \\
 &+&\frac{\lambda}{2}\sum_{\sigma}\sum_{i}
 V_i\rho^{\sigma}_{ii}
 + \frac{\lambda}{2}\sum_{\sigma}\sum^{i\neq j}_{ij}
 J_{ij}\rho^{\sigma}_{ij}
  \label{eq:I-6}
\end{eqnarray}
where the parameter $\lambda$ has been introduced again by the
replacement of $e^2$ by $\lambda e^2$ in $v(\vec{r}-\vec{r'})$. And
$V_i$ and $J_{ij}$ are
\begin{eqnarray}
V_i=-\int d^3r \phi^*_i(\vec{r})v(\vec{r})\phi_i(\vec{r}) = -e^2\int
d^3r \frac{|\phi_i(\vec{r})|^2}{r}
 \label{eq:Vi}
\end{eqnarray}
\begin{eqnarray}
J_{ij}=-\int d^3r \phi^*_i(\vec{r})v(\vec{r})\phi_j(\vec{r}).
 \label{eq:Jij}
\end{eqnarray}
Here the integral $V_i$ is the Coulomb interaction of a positive
charge ($+e$) at the position $\vec{r'}=0$ with the $\pi$ electron
at the position $\vec{r}$ near the site $i$. This positive charge is
in fact just the correlation hole moving together with the $\pi$
electron. The $\pi$ electron and the positive charge is at least a
lattice distant apart. The integral $J_{ij}$ ($i\neq j$) is a
two-center integral.

It can been seen from the definitions of them that off-diagonal
interactions refer to the overlaps of two $\pi$ wave functions lying
at two different sites[9] Thus they are actually very small compared
with the diagonal interactions $V_{ii,kk}$[17]. Thus, in ordinary
theoretical studies, only the diagonal interactions $V_{ii,kk}$ are
kept and the off-diagonal interactions are not considered. So only
the interactions $V_{ii,kk}$ in the first term of (\ref{eq:I-6}) are
kept in the physical study of low-dimensional $\pi$ electron
systems, just like that in the CDNO approximation. Similarly, the
integral $J_{ik}$ also refers to the overlap between two $\pi$
electronic wave functions respectively lying at two different
lattice sites and is a very small quantity compared with
$V_{ii,kk}$, then it can be omitted in some actual theoretical
calculations. It can be kept in theoretical calculations where more
accuracy is required.

By these considerations, (\ref{eq:I-6}) reduces into the form
\begin{eqnarray}
I \cong \frac{\lambda}{2}\sum_{\sigma\sigma'}\sum^{i\neq k}_{i,k}
  V_{ii,kk}\rho^{\sigma}_{ii}\rho^{\sigma'}_{kk}
 +\frac{\lambda}{2}\sum_{\sigma}\sum_{i}
 V_i\rho^{\sigma}_{ii}
  \label{eq:I-7}
\end{eqnarray}

Now we calculate $I_{HF}$. Because the procedure of calculating
$I_{HF}$ is given in the Ref.[1], here is not given out and readers
can be referred to that paper. Finally we have
\begin{eqnarray}
 I_{HF} &=& \lambda \sum_{\sigma,ik}^{i\neq k}V_{ii,kk}
 (\rho^{\sigma}_{ii} \rho^{\sigma}_{kk}
-\rho^{\sigma}_{ik}\rho^{\sigma}_{ki}).
 \label{eq:I-HF-4}
\end{eqnarray}

Inserting (\ref{eq:I-7}) and (\ref{eq:I-HF-4}) into (\ref{eq:Ec}),
the expression of $E_c$ is obtained
\begin{eqnarray}
 E_c &=&\frac{\lambda}{2}[-\frac{1}{2}\sum_{\sigma,i}
 |V_i|\rho^{\sigma}_{ii}
  + \sum_{\sigma,ik}^{i\neq k}V_{i,k}
 (\rho^{\sigma}_{ik})^2 ]
 \label{eq:Ec-2}
\end{eqnarray}
This is a more general formula for the electronic correlation
energy. The correlation energy can be acquired using this formula
when the averages of the diagonal charge density
$\rho^{\sigma}_{ii}$ and the off-diagonal charge density
$\rho^{\sigma}_{ik}$ ($i\neq k$) between the sites $i$ and $k$ and
also $V_{i,k}$ $(i\neq k)$ are known. When all $|V_i|$ is assumed to
take the nearest-neighbor Coulomb interaction $v$ and
$V_{i,k}=V_{ii,kk}=v$ with $k=i+1$, the above formula
(\ref{eq:Ec-2}) reduces to Eq.(15) obtained in the Ref.[1] which is
a special case of this formula.

When all kinds of electron-electron interactions are kept, the
correlation energy formula is given by
\begin{eqnarray}
E'_c &=&\frac{\lambda}{2}
 [-\frac{1}{2}\sum_{\sigma,i}
 |V_i|\rho^{\sigma}_{ii}
 + \frac{1}{2}\sum_{\sigma\sigma'}\sum^{i\neq k}_{i,k}
 V_{i,k}\rho^{\sigma}_{ik}\rho^{\sigma'}_{ik}
 \nonumber \\
&+& \frac{1}{2}\sum_{\sigma\sigma'}\sum^{k\neq l}_{i,kl}
 X_{ii,kl}\rho^{\sigma}_{ii}\rho^{\sigma'}_{kl}
+ \frac{1}{2}\sum_{\sigma\sigma'}\sum^{i\neq j,k\neq l}_{ij,kl}
 W_{ij,kl}\rho^{\sigma}_{ij}\rho^{\sigma'}_{kl}
 + \frac{1}{2}\sum_{\sigma}\sum^{i\neq j}_{ij}
 J_{ij}\rho^{\sigma}_{ij} ]
  \label{eq:Ec-3}
\end{eqnarray}
where $X_{ii,kl}\equiv V_{ii,kl}$ $(k\neq l)$, $W_{ij,kl}\equiv
V_{ij,kl}$ ($i\neq j,k\neq l$). Here the off-diagonal interaction
$X_{ii,kl}$ is called the bond-site interaction, the off-diagonal
interaction $W_{ij,kl}$ is called the bond-bond interaction. This is
general form of the correlation energy which includes all kinds of
the electron-electron interactions (the diagonal and the
off-diagonal interactions).

\section{Estimation of correlation energies}

For the situation where only the nearest-neighbor interactions are
considered, the above (\ref{eq:Ec-3}) can be simplified. For the
one-dimensional $\pi$ electron conjugated polymers with $N$ carbon
atoms such as the bond-alternated chain polyacetylene (PA), in the
bond order wave (BOW) phase, the average charge density
$\rho^{\sigma}_{ii}=1/2$, the bond charge density (also called the
bond-order matrix)
$\rho^{\sigma}_{ii+1}=\rho_{ii+1}=\bar\rho+(-1)^i\delta\rho$. Thus
we have the correlation energy formula from (\ref{eq:Ec-3})
\begin{eqnarray}
 E''_c &=&\frac{\lambda N}{2}
 [-\frac{v}{2} + 2(v+W)(\bar\rho^2+(\delta\rho)^2)
 + (X+J)\bar\rho ]
  \label{eq:Ec-5}
\end{eqnarray}
where $v\gg |X|>W>0$, $v\gg |J|$, and $X<0$, $J<0$. Note
(\ref{eq:Ec-5}) is suitable to an even number conjugated polymer
chain with single bond and double bond. For a odd number conjugated
polymer, there is a more term appearing in (\ref{eq:Ec-5}):
$-4(v+W)\bar\rho\delta\rho $ and $-(X+J)\delta\rho$.

The values of $v$,$W$ and $X$ have been already estimated before for
PA, which are shown in Table 1, and the value of J is not known.
However, we may estimated the value of $J$ through a rational
comparison between $X$ and $J$. According to the definitions of the
integrals of $X(<0)$ and $J(<0)$ in the nearest-neighbor situation,
$X$ is $\int \int d^3r
d^3r'\phi^*_i(\vec{r})\phi^*_i(\vec{r'})\times
v(\vec{r}-\vec{r'})\phi_i(\vec{r'})\phi_{i+1}(\vec{r})$, and $|J|$
is about $\int d^3r \phi^*_i(\vec{r})v(\vec{r})\phi_{i+1}(\vec{r})$,
where $v(\vec{r}-\vec{r'})$ is the Coulomb interaction between two
electronic charges, then it may be estimated that the maximum of
$|X|$ is about $\frac{2e^2}{a}\int
d^3r\phi^*_i(\vec{r})\phi_i(\vec{r})\int
d^3r'\phi^*_i(\vec{r'})\phi_{i+1}(\vec{r'})$, and the maximum of
$|J|$ is about $\frac{2e^2}{a}\int
d^3r'\phi^*_i(\vec{r'})\phi_{i+1}(\vec{r'})$, where $a$ is the
lattice constant and the wave functions $\phi_i(\vec{r})$ is
normalized. So the values of $|X|$ and $|J|$ are approximately
comparable with each other, that is, $|X|\sim |J|$. Therefore the
correlation energies formula for the nearest-neighbor interactions
reads
\begin{eqnarray}
  E'''_c &=&\frac{\lambda N}{2}
 [-\frac{v}{2} + 2(v+W)(\bar\rho^2+(\delta\rho)^2)
 + 2X\bar\rho ]
  \label{eq:Ec-6}
\end{eqnarray}
According to this formula we may estimate the correlation energies
under different off-diagonal interactions. In $ab$ $innito$
numerical computations for many-electronic systems of atoms,
molecules and solid state, the values of correlation energies are
mixed in exchange-correlation energies $E_{ex}$ instead of a
separate quantity, and moreover the off-diagonal interactions can
not be calculated concretely. Those are shortcomings of the $ab$
$inito$ numerical calculations. Through the formula here we may
clearly see the effect of the off-diagonal interactions on the
correlation energies and may discuss the physical properties such as
the band structure and electronic structure as well as other
physical properties related to elemental excitations. The results of
the calculations are shown in Table 2.

\begin{table}
 \centering
 \caption{The relations of the off-diagonal interactions and the
 screening factor $\beta$ when $U=6$ eV. The unit is eV.}
 \begin{tabular}{ccccc}
  \hline\hline
  $\beta=$ & 1 & 3 & 5 & 7 \\
   \hline
  $v=$ & 0.39U & 0.11U & 0.07U & 0.05U \\
  $X=$ & -0.0234U & -0.0198U & -0.0315U & -0.0385U \\
  $W=$ & 0.0078U & 0.011U & 0.0182U & 0.0215U \\
  \hline \hline\\
\end{tabular}
\end{table}

\begin{table}
 \centering
 \caption{The correlation energies $\epsilon_c$($\equiv E'''_c/N$,
 N is number of the $\pi$ electrons in a conjugated polymer chain)
 at different screening $\beta$ and different off-diagonal
 interactions $X$ and $W$ when $U=6$ eV. The unit is eV.}
 \begin{tabular}{cccccc}
  \hline\hline
   $\beta$ & 1 & 3 & 5 & 7  \\
   \hline
  $\epsilon_c(W=X=0)$ & -0.1673 & -0.0472 & -0.03 & -0.0215 \\
  $\epsilon_c(X\neq 0, W=0)$ & -0.1894 & -0.0659 & -0.0597 & -0.0578 \\
  $\epsilon_c(X=0,W\neq 0)$ & -0.1648 & -0.0437 & -0.0242 & -0.0145 \\
  $\epsilon_c(X\neq 0,W\neq 0)$ & -0.1869 & -0.0623 & -0.0539 & -0.0509 \\
  \hline \hline\\
\end{tabular}
\end{table}

\section{Discussion and summary}

The integral $V_i$ is an integral of a positive charge (Correlation
hole) at the origin with all other electrons in the whole system.
Although the form of the integral $V_i$ is simple but its
calculation still is not direct, and the approximate calculation
about it is available for some special systems. The simplest is that
$V_i$ takes the nearest-neighbor interaction $v$. The more
complicated calculation is left in the later study and here we do
not discuss its computation. Similarly direct computation of the
integral $J$ is not considered here and is left in the later study.
The site charge density $\rho^{\sigma}_{ii}$ (that is, the diagonal
elements of charge density) and the bond charge density
$\rho^{\sigma}_{ik}$ ($i\neq k$) (that is, the off-diagonal parts of
charge density) can be calculated directly from the elliptic
integrals for a rigid conjugated polymers[9], and also can be drawn
from the results of numerical computation including $ab$ $initio$
computations.

It is all known that the values of the off-diagonal interactions are
related to the screening values[3]. Table 1 lists the values of the
diagonal interaction $v$ and the off-diagonal interactions $X$ and
$W$ according to the computing results of the Ref.[3]. According to
the Ref.[3], the order of the diagonal and the off-diagonal
interactions are $v>|X|>W$. All interactions (the diagonal and the
off-diagonal interactions ) are related to the on-site Hubbard
interaction which was discussed in the Ref.[9]. When $U$ is given,
the values of $v$, $W$, and $X$ are determined. In our calculation,
$U=6$. It is seen from the Table 1 that at the normal screening case
($\beta\sim 1$) and the middle screening case ($\beta\sim 3$), $W$
is much smaller than $X$, and $X$ is much smaller than $v$. Only
with increasing the screening up to the case $\beta \ge 5$, both $W$
and $X$ become stronger, they even go over the value of $v$. Anyway,
at the normal screening and the middle screening, the off-diagonal
interactions $W$ and $X$ may be regarded as a perturbation relative
to $v$. Interesting is that the off-diagonal interaction $X$ is
negative, which may be thought to be related to the possibility of
the superconductivity of the conjugated polymers[12].

It is seen from our calculating results shown in Table 2 that when
the screening $\beta$ increases from normal to strong, the
correlation energies diminish from large to small whatever with the
off-diagonal interactions ($X$, $W$) and without the off-diagonal
interaction ($X$, $W$). It is seen from Table 2 that the
off-diagonal interaction $X$ increases the correlation energies at
most and the off-diagonal interaction $W$ decreases the correlation
energies compared with those without considering the off-diagonal
interactions $X$ or $W$. When both $X$ and $W$ are considered in
calculations, the correlation energies increase compared with those
without considering both $X$ and $W$. It has been known that the
bond-site interaction $X$ was used to discuss the superconductivity
of the polymers[12][17]. Here the $X$ is shown to have contribution
to the correlation energy, which may imply important significance of
the long-range correlation to the superconductivity of the polymers.

It has been known that under fixed $U$, the exciton binding energy
decreases with increasing the screening $\beta$[9]. Here under fixed
$U$, the correlation energies decrease with increasing the screening
$\beta$. About the screening $\beta$, the bond density correlation
of the $\pi$ electrons, the off-diagonal interactions, the exciton
bonding energy, and their relations has been detailed in the
Ref.[9]. $\beta$ is a cutoff parameter of the Coulomb interaction.
The larger the $\beta$, the weaker the Coulomb interaction. Thus the
factor $\beta$ reflects a kind of screening effect on the $\pi$
electrons in the conjugated polymers. But the larger the $\beta$,
the changes of the relative values among $v$, $W$, and $X$, see
Table 1. This shows the significance of the screening $\beta$: the
larger $\beta$, the stronger the bond density correlations. Notice
here that we should not confuse the correlation energy with the bond
density correlation. Therefore, we see that with increasing the
screening $\beta$ ($\ge 3$), the off-diagonal interaction $X$
increases much, the exciton binding energy decreases, the bond
density correlation increases, and the correlation energies also
increase. It is found from the Table 2 that the correlation energies
decrease rapidly with increasing the screening when $\beta$ is from
1 to 7 when without considering the off-diagonal interactions $W$
and $X$. But when considering only $X$ or considering both $X$ and
$W$, the change of the correlation energies with increasing
screening ($\beta\ge 3$) is not so big, this is because when the
screening $\beta$ increases the off-diagonal interaction $X$ also
increases very much.

In summary, in this paper the general formula of the correlation
energies has been deduced with the off-diagonal interactions $X$ and
$W$ by the first time. The effect of the off-diagonal interactions
$X$ and $W$ on the correlation energies are discussed. It is found
that the effects of the off-diagonal interactions $W$ and $X$ on the
correlation energies are opposite. Given the screening $\beta$, the
off-diagonal interaction $X$ increase the correlation energies and
the off-diagonal interaction $W$ decreases the correlation energies.
But the influence of $X$ on the correlation energies is much larger
than that of $W$ on the correlation energies, and the correlation
energies decrease with increasing the screening factor $\beta$.

\begin{center}
\large {\bf ACKNOWLEDGMENT}
\end{center}
Support from the 985 Scientific Research Foundation of ChongQing
University, ChongQing, P.R. China is acknowledged.


\end{document}